\documentclass[english,superscriptaddress,showpacs]{revtex4}
\usepackage{graphicx}
\usepackage{amssymb}

\usepackage{psfrag}

\usepackage{babel}

\begin{document}

\title{A general derivation of differential cross section in quark-quark
scatterings at fixed impact parameter}

\author{Shou-wan Chen}

\affiliation{Department of Modern Physics and Interdisciplinary Center for Theoretical Study, 
University of Science and Technology of China, Hefei, Anhui 230026, People's Republic of China}

\author{Jian Deng}

\affiliation{Department of Modern Physics and Interdisciplinary Center for Theoretical Study, 
University of Science and Technology of China, Hefei, Anhui 230026, People's Republic of China}

\author{Jian-hua Gao}

\affiliation{Department of Physics, Shandong University, Jinan, Shandong 250100,
People's Republic of China}

\author{Qun Wang}

\affiliation{Department of Modern Physics and Interdisciplinary Center for Theoretical Study, 
University of Science and Technology of China, Hefei, Anhui 230026, People's Republic of China}

\begin{abstract}
We propose a general derivation of differential cross section in quark-quark
scatterings at fixed impact parameters. The derivation is well defined
and free of ambiguity in the conventional one. The approach can be
applied to a variety of partonic and hadronic scatterings in low or
high energy particle collisions. 
\end{abstract}


\maketitle

\section{Introduction}

The global polarization effect in quark scatterings in non-central
heavy-ion collisions has been predicted due to orbital angular momenta
resided in the system as a result of longitudinal flow shear \cite{Liang:2004ph,Voloshin:2004ha,Betz:2007kg,Gao:2007bc}.
The effect is defined with respect to the direction orthogonal to
the reaction plane determined by the vector of impact-parameter and
the beam momentum. The polarization of quarks can be partially carried
by hadrons containing these quarks via hadronization. For example,
as a consequence of this global polarization, vector mesons can have
spin alignments in non-central heavy-ion collisions \cite{Liang:2004xn}.
Polarized photons are recently proposed as a good probe to the polarized
quarks \cite{Ipp:2007ng}. 

In Ref. \cite{Gao:2007bc}, the polarization in the quark-quark scatterings
has been estimated in a hot medium. The differential cross-section
with respect to the partonic impact parameter is derived by inserting
into the differential cross-section in mementum space a delta function
for transverse momenta, which can in turn be written as an integral
over the partonic impact parameter. This derivation has an ambiguity.
In this note we will give an alternative and a general way of deriving
the differential cross-section with respect to the partonic impact
parameter. We will show that the differential cross section with respect
to impact parameter in our framework is well defined and free of ambiguity.

\section{Kinematics setup }

The geometry of a nucleus-nucleus collision at impact parameter $b$
is illustrated in Fig. 1 of Ref. \cite{Liang:2004ph}. We assume
that two nuclei move along $\pm z$ directions and collide at $z=0$.
The reaction plane is in the $y$ direction. We consider the polarization
along the $y$ direction. 

Let us consider the scattering of quarks with different flavors $(P_{1},\lambda_{1})+(P_{2},\lambda_{2})\rightarrow(P_{3},\lambda_{3})+(P_{4},\lambda_{4})$
through the Hard-Thermal-Loop (HTL) resummed gluon propagators, where
$\lambda_{i}$ with $(i=1,2,3,4)$ are spin states and $P_{i}=(E_{i},p_{iz},\mathbf{p}_{iT})$
are 4-momenta for colliding quarks, with longitudinal and transverse
momenta $p_{iz}$ and $\mathbf{p}_{iT}$ and energies $E_{i}=\sqrt{p_{iz}^{2}+p_{iT}^{2}}$.
Note that we treat all quarks massless. We assume the initial momenta
are along the $z$ direction, i.e. $p_{1z}=-p_{2z}>0$ and $\mathbf{p}_{1T}=\mathbf{p}_{2T}=0$.
We will use the shorthand notation $\mathbf{p}_{T}\equiv\mathbf{p}_{3T}$.
The total energy in the center-of-mass frame is denoted by $\sqrt{s}=\sqrt{(P_{1}+P_{2})^{2}}$.

\section{Ambiguity of differential cross section with respect to impact parameter
in conventional treatment}

In a conventional derivation of the differential cross section at
impact parameter $\mathbf{x}_{T}$, one starts with the one in momentum
space, \begin{eqnarray}
d\sigma & = & \frac{1}{16|P_{1}\cdot P_{2}|}\sum_{\lambda_{1},\lambda_{2},\lambda_{4}}|M(P_{1}\lambda_{1},P_{2}\lambda_{2},P_{3}\lambda_{3},P_{4}\lambda_{4})|^{2}\nonumber \\
 &  & \times(2\pi)^{4}\delta^{(4)}(P_{1}+P_{2}-P_{3}-P_{4})\frac{d^{3}\mathbf{p}_{3}}{(2\pi)^{3}2E_{3}}\frac{d^{3}\mathbf{p}_{4}}{(2\pi)^{3}2E_{4}},\label{eq:cross-section-m}\end{eqnarray}
where a sum over the spin states $\lambda_{1},\lambda_{2},\lambda_{4}$
except $\lambda_{3}$ and an average (with a factor $1/4$) over the
initial spin states have been taken. One can integrate out $\mathbf{p}_{4}$
by consuming the delta function $\delta^{(3)}(\mathbf{p}_{1}+\mathbf{p}_{2}-\mathbf{p}_{3}-\mathbf{p}_{4})$
for 3-momentum conservation, and then carry out the integral over
$p_{3z}$ to remove $\delta(E_{1}+E_{2}-E_{3}-E_{4})$ for the energy
conservation which gives a factor \begin{eqnarray}
\int dp_{3z}\delta(E_{1}+E_{2}-E_{3}-E_{4}) & = & \sum_{i=1,2}\frac{E_{3}E_{4}}{\left|E_{3}p_{4z}^{(i)}-E_{4}p_{3z}^{(i)}\right|},\label{eq:approx2}\end{eqnarray}
where $p_{4z}^{(i)}=p_{1z}+p_{2z}-p_{3z}^{(i)}$ and $p_{3z}^{(i)}$
are roots of the energy conservation equation and given by \begin{eqnarray}
p_{3z}^{(1/2)} & = & \pm\frac{E_{1}+E_{2}}{2}\left[1-\frac{2p_{T}^{2}}{|p_{1z}p_{2z}|-p_{1z}p_{2z}}\right]^{1/2}+\frac{p_{1z}+p_{2z}}{2}.\label{eq:p3z-root}\end{eqnarray}
Then Eq. (\ref{eq:cross-section-m}) is simplified as \begin{eqnarray}
d\sigma & = & \frac{1}{64|P_{1}\cdot P_{2}|}\sum_{i=1,2}\sum_{\lambda_{1},\lambda_{2},\lambda_{4}}|M(P_{3}\lambda_{3},P_{4}\lambda_{4})|^{2}\frac{1}{\left|E_{3}p_{4z}^{(i)}-E_{4}p_{3z}^{(i)}\right|}\frac{d^{2}\mathbf{p}_{T}}{(2\pi)^{2}},\end{eqnarray}
where we have suppressed the $(P_{1}\lambda_{1},P_{2}\lambda_{2})$
dependence of the amplitude. Note that $\mathbf{p}_{4T}=-\mathbf{p}_{3T}=-\mathbf{p}_{T}$
is implied in the above expression due to momentum conservation and
the assumption that the initial state momenta are along the $z$-axis,
$\mathbf{p}_{1T}=\mathbf{p}_{2T}=0$. One can rewrite it by inserting
a delta function for transverse momenta, \begin{eqnarray}
d\sigma & = & \frac{1}{64|P_{1}\cdot P_{2}|}\int\frac{d^{2}\mathbf{p}_{T}}{(2\pi)^{2}}\frac{d^{2}\mathbf{p}'_{T}}{(2\pi)^{2}}(2\pi)^{2}\delta^{(2)}(\mathbf{p}_{T}-\mathbf{p}'_{T})\nonumber \\
 &  & \times\sum_{i=1,2}\sum_{\lambda_{1},\lambda_{2},\lambda_{4}}\frac{1}{\left|E_{3}p_{4z}^{(i)}-E_{4}p_{3z}^{(i)}\right|}M(P_{3}\lambda_{3},P_{4}\lambda_{4})M^{*}(P'_{3}\lambda_{3},P'_{4}\lambda_{4})\nonumber \\
 & = & \frac{1}{64|P_{1}\cdot P_{2}|}\int d^{2}\mathbf{x}_{T}\frac{d^{2}\mathbf{p}_{T}}{(2\pi)^{2}}\frac{d^{2}\mathbf{p}'_{T}}{(2\pi)^{2}}e^{i(\mathbf{p}_{T}-\mathbf{p}'_{T})\cdot\mathbf{x}_{T}}\nonumber \\
 &  & \times\sum_{i=1,2}\sum_{\lambda_{1},\lambda_{2},\lambda_{4}}\frac{1}{\left|E_{3}p_{4z}^{(i)}-E_{4}p_{3z}^{(i)}\right|}M(P_{3}\lambda_{3},P_{4}\lambda_{4})M^{*}(P'_{3}\lambda_{3},P'_{4}\lambda_{4}),\label{eq:diff-crs-xt1}\end{eqnarray}
where $P'_{3}=(E'_{3},p_{3z},\mathbf{p}'_{T})$ and $P'_{4}=(E'_{4},p_{4z},-\mathbf{p}'_{T})$.
Then we obtain the differential cross section at impact parameter
$\mathbf{x}_{T}$, \begin{eqnarray}
\frac{d^{2}\sigma}{d^{2}\mathbf{x}_{T}} & = & \frac{1}{64|P_{1}\cdot P_{2}|}\int\frac{d^{2}\mathbf{p}_{T}}{(2\pi)^{2}}\frac{d^{2}\mathbf{p}'_{T}}{(2\pi)^{2}}e^{i(\mathbf{p}_{T}-\mathbf{p}'_{T})\cdot\mathbf{x}_{T}}\nonumber \\
 &  & \times\sum_{i=1,2}\sum_{\lambda_{1},\lambda_{2},\lambda_{4}}\frac{1}{\left|E_{3}p_{4z}^{(i)}-E_{4}p_{3z}^{(i)}\right|}M(P_{3}\lambda_{3},P_{4}\lambda_{4})M^{*}(P'_{3}\lambda_{3},P'_{4}\lambda_{4}).\label{eq:diff-crs-xt2}\end{eqnarray}
Note that the above expression of $d^{2}\sigma/d^{2}\mathbf{x}_{T}$
is not unique, since one can make the replacement in Eq. (\ref{eq:diff-crs-xt1}),
for example, \begin{eqnarray}
\frac{1}{\left|E_{3}p_{4z}^{(i)}-E_{4}p_{3z}^{(i)}\right|} & \rightarrow & \frac{1}{\left|E_{3}p_{4z}^{(i)}-E_{4}p_{3z}^{(i)}\right|^{a_{1}}}\frac{1}{\left|E'_{3}{p'}_{4z}^{(i)}-E'_{4}{p'}_{3z}^{(i)}\right|^{a_{2}}}\end{eqnarray}
with $a_{1}+a_{2}=1$, while keeping the total cross section unchanged.
Of course one can make many other choices which conserve the total
cross section. In Ref. \cite{Gao:2007bc}, $a_{1}=a_{2}=1/2$ is
used, while Eq. (\ref{eq:diff-crs-xt2}) implies $a_{1}=1,a_{2}=0$.
So we see that there is an ambiguity in $d^{2}\sigma/d^{2}\mathbf{x}_{T}$,
or in other word, $d^{2}\sigma/d^{2}\mathbf{x}_{T}$ is not unique
by this definition. The problem arises when one calculates quantities
like the polarization when the integral over $\mathbf{x}_{T}$ is
not made in the whole space (therefore the delta function $\delta^{(2)}(\mathbf{p}_{T}-\mathbf{p}'_{T})$
is not recovered), which would lead to different or inconsistent results.

\section{Differential cross section at fixed impact parameter}

In order to solve the ambiguity in the previous section, we will derive
in this section the cross sections of parton-parton scatterings at
fixed impact parameter in a general approach. To this end, we need
to introduce particle states labeled by transverse positions and longitudinal
momenta, which we call states in the mixed representation. They are
connected with states in momentum space by Fourier transform in transverse
sector. We express in the mixed representation the final states in
the scatterings as \begin{eqnarray}
\left|p_{3z},\lambda_{3},\mathbf{x}_{3T}\right\rangle  & = & \int\frac{A_{T}d^{2}\mathbf{p}_{3T}}{(2\pi)^{2}}e^{i\mathbf{p}_{3T}\cdot\mathbf{x}_{3T}}\left|\mathbf{p}_{3},\lambda_{3}\right\rangle ,\nonumber \\
\left|p_{4z},\lambda_{4},\mathbf{x}_{4T}\right\rangle  & = & \int\frac{A_{T}d^{2}\mathbf{p}_{4T}}{(2\pi)^{2}}e^{i\mathbf{p}_{4T}\cdot\mathbf{x}_{4T}}\left|\mathbf{p}_{4},\lambda_{4}\right\rangle ,\end{eqnarray}
where $A_{T}$ is the area in the transverse plane. The S-matrix element
from the initial to final states then reads \begin{eqnarray}
S_{fi} & = & \left\langle p_{3z},\lambda_{3},\mathbf{x}_{3T};p_{4z},\lambda_{4},\mathbf{x}_{4T}\right|S\left|\mathbf{p}_{1},\lambda_{1},\mathbf{p}_{2},\lambda_{2}\right\rangle \nonumber \\
 & = & \int\frac{A_{T}d^{2}\mathbf{p}_{3T}}{(2\pi)^{2}}\frac{A_{T}d^{2}\mathbf{p}_{4T}}{(2\pi)^{2}}e^{-i\mathbf{p}_{3T}\cdot\mathbf{x}_{3T}}e^{-i\mathbf{p}_{4T}\cdot\mathbf{x}_{4T}}(2\pi)^{4}\delta^{(4)}(P_{1}+P_{2}-P_{3}-P_{4})\nonumber \\
 &  & \times\frac{1}{\sqrt{2E_{1}V}}\frac{1}{\sqrt{2E_{2}V}}\frac{1}{\sqrt{2E_{3}V}}\frac{1}{\sqrt{2E_{4}V}}M(P_{3}\lambda_{3},P_{4}\lambda_{4}),\end{eqnarray}
where the final state momenta are $P_{3}=(E_{3},p_{3z},\mathbf{p}_{3T})$
and $P_{4}=(E_{4},p_{4z},\mathbf{p}_{4T})$. The squared matrix element
becomes \begin{eqnarray}
\left|S_{fi}\right|^{2} & = & \int\frac{A_{T}d^{2}\mathbf{p}_{3T}}{(2\pi)^{2}}\frac{A_{T}d^{2}\mathbf{p}_{4T}}{(2\pi)^{2}}\frac{A_{T}d^{2}\mathbf{p}'_{3T}}{(2\pi)^{2}}\frac{A_{T}d^{2}\mathbf{p}'_{4T}}{(2\pi)^{2}}e^{-i(\mathbf{p}_{3T}-\mathbf{p}'_{3T})\cdot\mathbf{x}_{3T}}e^{-i(\mathbf{p}_{4T}-\mathbf{p}'_{4T})\cdot\mathbf{x}_{4T}}\nonumber \\
 &  & \times(2\pi)^{8}\delta^{(4)}(P_{1}+P_{2}-P_{3}-P_{4})\delta^{(4)}(P_{1}+P_{2}-P'_{3}-P'_{4})\nonumber \\
 &  & \times\frac{1}{V^{4}}\frac{1}{16E_{1}E_{2}\sqrt{E_{3}E_{4}E'_{3}E'_{4}}}M(P_{3}\lambda_{3},P_{4}\lambda_{4})M^{*}(P'_{3}\lambda_{3},P'_{4}\lambda_{4}),\label{eq:sfi}\end{eqnarray}
where $P'_{3}=(E'_{3},p_{3z},\mathbf{p}'_{3T})$ and $P'_{4}=(E'_{4},p_{4z},\mathbf{p}'_{4T})$.
Note that the longitudinal momenta of $P_{3}$ and $P'_{3}$ are the
same, so are $P_{4}$ and $P'_{4}$. The transverse parts of the delta-functions
ensure $\mathbf{p}_{3T}=-\mathbf{p}_{4T}=\mathbf{p}_{T}$ and $\mathbf{p}'_{3T}=-\mathbf{p}'_{4T}=\mathbf{p}'_{T}$.
We can integrate out $\mathbf{p}'_{T}$ and $\mathbf{p}_{T}$ to get
rid of the four delta functions in the transverse sector and arrive
at \begin{eqnarray}
\left|S_{fi}\right|^{2} & = & A_{T}^{4}\int d^{2}\mathbf{p}_{T}d^{2}\mathbf{p}'_{T}e^{-i(\mathbf{p}_{T}-\mathbf{p}'_{T})\cdot(\mathbf{x}_{3T}-\mathbf{x}_{4T})}\delta^{(0,z)}(P_{1}+P_{2}-P_{3}-P_{4})\delta^{(0,z)}(P_{1}+P_{2}-P'_{3}-P'_{4})\nonumber \\
 &  & \times\frac{1}{V^{4}}\frac{1}{16E_{1}E_{2}\sqrt{E_{3}E_{4}E'_{3}E'_{4}}}M(P_{3}\lambda_{3},P_{4}\lambda_{4})M^{*}(P'_{3}\lambda_{3},P'_{4}\lambda_{4}),\end{eqnarray}
where $\delta^{(0,z)}$ denote the delta functions for the energy
and the $z$ component of momenta. The differential cross section
is then \begin{eqnarray}
d\sigma & = & \frac{d^{2}\mathbf{x}_{T}}{A_{T}}\frac{V}{4v_{rel}\tau}\sum_{\lambda_{1},\lambda_{2},\lambda_{4}}\int\frac{Ldp_{3z}}{2\pi}\frac{Ldp_{4z}}{2\pi}\left|S_{fi}\right|^{2}\nonumber \\
 & = & d^{2}\mathbf{x}_{T}\frac{1}{\tau}\frac{1}{64(2\pi)^{3}}\int d^{2}\mathbf{p}_{T}dp_{3z}d^{2}\mathbf{p}'_{T}dp_{4z}e^{-i(\mathbf{p}_{3T}-\mathbf{p}'_{3T})\cdot\mathbf{x}_{T}}\nonumber \\
 &  & \times\frac{1}{v_{rel}E_{1}E_{2}\sqrt{E_{3}E_{4}E'_{3}E'_{4}}}\delta^{(0,z)}(P_{1}+P_{2}-P_{3}-P_{4})\delta(E_{1}+E_{2}-E'_{3}-E'_{4})\nonumber \\
 &  & \times\sum_{\lambda_{1},\lambda_{2},\lambda_{4}}M(P_{3}\lambda_{3},P_{4}\lambda_{4})M^{*}(P'_{3}\lambda_{3},P'_{4}\lambda_{4}),\label{eq:diff-sigma}\end{eqnarray}
where we have defined $\mathbf{x}_{T}\equiv\mathbf{x}_{3T}-\mathbf{x}_{4T}$.
In the first line we have used that the differential cross section
is proportional to the fraction $d^{2}\mathbf{x}_{T}/A_{T}$ with
$A_{T}=\int d^{2}\mathbf{x}_{T}$. Here $L$ is the length along the
$z$ direction and $v_{rel}=|P_{1}\cdot P_{2}|/(E_{1}E_{2})$ the
relative velocity of incident partons. We also used $V=A_{T}L$ and
$2\pi\delta^{(z)}(0)=L$ (because two delta functions for $z$-momenta
are identical). Note that $\tau$ is the time period of the scattering.
It is obvious to see $d^{2}\sigma/d^{2}\mathbf{x}_{T}>0$ from Eq.
(\ref{eq:diff-sigma}). 

\begin{figure}
\caption{\label{fig:exact-result}(color online) The polarized and unpolarized
differential cross sections. The exact results from Eq. (\ref{eq:cross-section1})
is in the red solid line (unpolarized) and blue dashed line (polarized
part multiplied by a factor of 10). The sum of the unpolarized part
$d^{2}(\sigma_{upol}^{(0)}+\sigma_{upol}^{(1)})/d^{2}\mathbf{x}_{T}$
from Eqs. (\ref{eq:zero-order},\ref{eq:first-order1},\ref{eq:first-order2})
is in the red dotted line, while that of the polarized part $d^{2}(\sigma_{pol}^{(0)}+\sigma_{pol}^{(1)})/d^{2}\mathbf{x}_{T}$
(multiplied by a factor of 10) is in the blue dash-dotted line. The
parameters are chosen to be $\sqrt{s}=20$ GeV, $T=200$ MeV, $\tau=1.1$
fm and $\alpha_{s}=1$. In the perturbation approach to obtain $d^{2}(\sigma^{(0)}+\sigma^{(1)})/d^{2}\mathbf{x}_{T}$
we use $p_{t}^{cut}=3$ GeV. }

\includegraphics[scale=0.6]{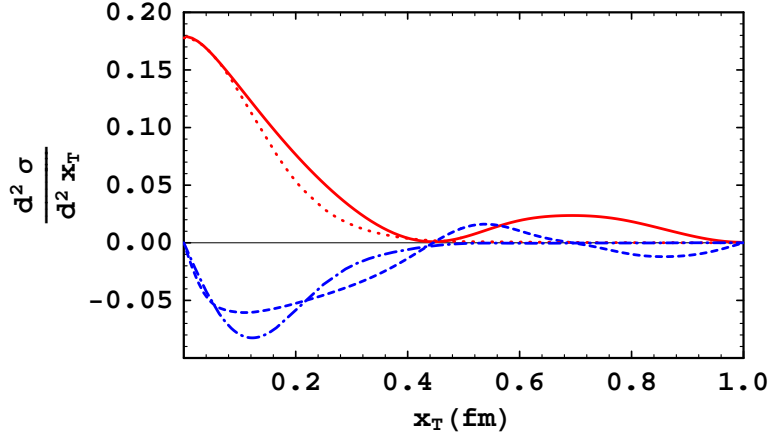}
\end{figure}

We can evaluate Eq. (\ref{eq:diff-sigma}) as follows. First we intergrate
out $p_{4z}$ to remove $\delta(p_{1z}+p_{2z}-p_{3z}-p_{4z})$. The
remaining two delta functions enforce energy conservation, which can
be removed by carrying out integrals over the magnitudes of transverse
momenta $p'_{T}$ and $p_{T}$. We end up with \begin{eqnarray}
\frac{d^{2}\sigma}{d^{2}\mathbf{x}_{T}} & = & \frac{1}{\tau}\frac{1}{64(2\pi)^{3}|P_{1}\cdot P_{2}|}\int_{0}^{2\pi}d\varphi\int_{0}^{2\pi}d\varphi'\int_{p_{3z}^{min}}^{p_{3z}^{max}}dp_{3z}e^{-ip_{T}(\cos\varphi-\cos\varphi')x_{T}}\frac{E_{3}E_{4}}{(E_{3}+E_{4})^{2}}\nonumber \\
 &  & \times\sum_{\lambda_{1},\lambda_{2},\lambda_{4}}M(P_{3}\lambda_{3},P_{4}\lambda_{4})M^{*}(P'_{3}\lambda_{3},P'_{4}\lambda_{4}),\label{eq:cross-section1}\end{eqnarray}
where $\varphi$ and $\varphi'$ are azimuthal angles of $\mathbf{p}_{T}$
and $\mathbf{p}'_{T}$ relative to the direction of $\mathbf{x}_{T}$
respectively. The magnitude of transverse momentum satisfying the
energy conservation for $P_{3,4}$ and $P'_{3,4}$ is proved to be
the same and given by \begin{eqnarray}
p_{T} & = & \left\{ \left[\frac{(E_{1}+E_{2})^{2}+p_{3z}^{2}-(p_{1z}+p_{2z}-p_{3z})^{2}}{2(E_{1}+E_{2})}\right]^{2}-p_{3z}^{2}\right\} ^{1/2}.\label{eq:p3t0}\end{eqnarray}
The energies are \begin{eqnarray}
E_{3}=E'_{3} & = & \sqrt{p_{3z}^{2}+p_{T}^{2}},\nonumber \\
E_{4}=E'_{4} & = & \sqrt{(p_{1z}+p_{2z}-p_{3z})^{2}+p_{T}^{2}}.\end{eqnarray}
The integral $p_{3z}$ is in the range $[p_{3z}^{min},p_{3z}^{max}]$
where \begin{eqnarray}
p_{3z}^{max} & = & \frac{(E_{1}+E_{2})^{2}-(p_{1z}+p_{2z})^{2}}{2(E_{1}+E_{2})}\left[1-\frac{p_{1z}+p_{2z}}{E_{1}+E_{2}}\right]^{-1},\nonumber \\
p_{3z}^{min} & = & -\frac{(E_{1}+E_{2})^{2}-(p_{1z}+p_{2z})^{2}}{2(E_{1}+E_{2})}\left[1+\frac{p_{1z}+p_{2z}}{E_{1}+E_{2}}\right]^{-1}.\end{eqnarray}
The amplitude square is given by \begin{eqnarray}
M(P_{3}\lambda_{3},P_{4}\lambda_{4})M^{*}(P'_{3}\lambda_{3},P'_{4}\lambda_{4}) & = & g^{4}c_{qq}E_{1}E_{2}\sqrt{E_{3}E_{4}E'_{3}E'_{4}}J_{1}^{\mu\mu'}J_{2}^{\nu\nu'}\Delta_{\mu\nu}(P_{1}-P_{3})\Delta_{\mu'\nu'}^{*}(P_{1}-P_{3'})\label{eq:mm-s02}\end{eqnarray}
Here $g$ is the quark-gluon coupling constant and $\alpha_{s}=g^{2}/(4\pi)$.
The color factor for $qq$ scatterings is $c_{qq}=(1/N_{c}^{2})(\delta^{ab}\delta^{ab}/4)=2/9$,
where $1/N_{c}^{2}$ is from the average over the initial state colors.
$J_{1}^{\mu\mu'}$ and $J_{2}^{\nu\nu'}$ are tensors only dependent
on momentum directions, \begin{eqnarray}
J_{1}^{\mu\mu'} & \equiv & \frac{1}{E_{1}\sqrt{E_{3}E'_{3}}}\mathrm{Tr}[u(P'_{3},\lambda_{3})\overline{u}(P_{3},\lambda_{3})\gamma^{\mu}P_{1\sigma}\gamma^{\sigma}\gamma^{\mu'}]\nonumber \\
J_{2}^{\nu\nu'} & \equiv & \frac{1}{E_{2}\sqrt{E_{4}E'_{4}}}\sum_{\lambda_{4}}\mathrm{Tr}[u(P'_{4},\lambda_{4})\overline{u}(P_{4},\lambda_{4})\gamma^{\nu}P_{2\sigma}\gamma^{\sigma}\gamma^{\nu'}]\end{eqnarray}
where $u(P_{i},\lambda_{i})$ denotes the spinor for the parton $i$
with the spin state $\lambda_{i}$ along the reference direction $\mathbf{n}\equiv\mathbf{e}_{y}$,
and $\overline{u}_{i,\lambda_{i}}=u^{\dagger}(P_{i},\lambda_{i})\gamma_{0}$
is its conjugate. $\Delta_{\mu\nu}$ is the HTL resummed gluon propagator
\cite{Weldon:1982aq,Braaten:1989mz,Heiselberg:1996xg}. Here we only
take the magnetic gluon exchange into account which invloves the magnetic
mass $\mu_{m}$ is introduced to regulate the divergence arising from
the soft gluon exchange. The magnetic mass $\mu_{m}$ is proportional
to the temperature $T$, $\mu_{m}=0.255(N_{c}/2)^{1/2}g^{2}T$. The
numerical result from Eq. (\ref{eq:cross-section1}) is given in Fig.
\ref{fig:exact-result} for collisions in the center-of-mass system
of the colliding quarks. The polarized part is much less than unpolarized
one. Both the unpolarized and polarized differential cross sections
show an oscillation feature. 

Thers is an alternative way to do the integrals in Eq. (\ref{eq:diff-sigma}).
One can intergrate out $p_{4z}$ to remove $\delta(p_{1z}+p_{2z}-p_{3z}-p_{4z})$
and then $p_{3z}$ to remove $\delta(E_{1}+E_{2}-E_{3}-E_{4})$ as
in Eq. (\ref{eq:approx2}). We finally obtain the differential cross
section with respect to the impact parameter, \begin{eqnarray}
\frac{d^{2}\sigma}{d^{2}\mathbf{x}_{T}} & = & \frac{1}{\tau}\frac{1}{64(2\pi)^{3}}\int d^{2}\mathbf{p}_{T}d^{2}\mathbf{p}'_{T}e^{-i(\mathbf{p}_{T}-\mathbf{p}'_{T})\cdot\mathbf{x}_{T}}\delta(E_{3}+E_{4}-E'_{3}-E'_{4})\nonumber \\
 &  & \times\sum_{i=1,2}\sum_{\lambda_{1},\lambda_{2},\lambda_{4}}\frac{1}{|P_{1}\cdot P_{2}|\sqrt{E_{3}E_{4}E'_{3}E'_{4}}}\frac{E_{3}E_{4}}{\left|E_{3}p_{4z}^{(i)}-E_{4}p_{3z}^{(i)}\right|}\nonumber \\
 &  & \times M(P_{3}\lambda_{3},P_{4}\lambda_{4})M^{*}(P'_{3}\lambda_{3},P'_{4}\lambda_{4}),\label{eq:approx3}\end{eqnarray}
Note that all final state energies are functions of $\mathbf{p}_{T}$
and $\mathbf{p}'_{T}$. The root $p_{3z}^{(i)}$ is given by Eq. (\ref{eq:p3z-root})
and $p_{4z}^{(i)}$ is given by $p_{4z}^{(i)}=p_{1z}+p_{2z}-p_{3z}^{(i)}$.
We now rewrite the remaining delta-function into an integral over
a time $t$ which is the conjugate variable of the uncertainty of
the final state energies arising from the specified transverse positions
in the final state, \begin{eqnarray}
2\pi\delta(E_{3}+E_{4}-E'_{3}-E'_{4}) & \approx & \int_{-\tau/2}^{\tau/2}dt\; e^{i(E_{3}+E_{4}-E'_{3}-E'_{4})t},\end{eqnarray}
where we have assumed $T\rightarrow\infty$. Inserting the above into
Eq. (\ref{eq:approx3}), we obtain \begin{eqnarray}
\frac{d^{2}\sigma}{d^{2}\mathbf{x}_{T}} & = & \frac{1}{64(2\pi)^{4}|P_{1}\cdot P_{2}|}\int d^{2}\mathbf{p}_{T}d^{2}\mathbf{p}'_{T}\left\langle e^{-i(\mathbf{p}_{T}-\mathbf{p}'_{T})\cdot\mathbf{x}_{T}}\right.\nonumber \\
 &  & \times\sum_{i=1,2}\sum_{\lambda_{1},\lambda_{2},\lambda_{4}}\frac{1}{\sqrt{E_{3}E_{4}E'_{3}E'_{4}}}\frac{E_{3}E_{4}}{\left|E_{3}p_{4z}^{(i)}-E_{4}p_{3z}^{(i)}\right|}\nonumber \\
 &  & \left.\times M(P_{3}\lambda_{3},P_{4}\lambda_{4})M^{*}(P'_{3}\lambda_{3},P'_{4}\lambda_{4})\right\rangle _{t},\label{eq:approx4}\end{eqnarray}
where the time average $\left\langle \cdots\right\rangle _{t}$ is
defined by \begin{equation}
\left\langle \cdots\right\rangle _{t}\equiv\frac{1}{\tau}\int_{-\tau/2}^{\tau/2}dt\;(\cdots)\, e^{i(E_{3}+E_{4}-E'_{3}-E'_{4})t}.\end{equation}
 At small transverse momenta in the center-of-mass system where $p_{T}\sim p'_{T}\ll\sqrt{s}$,
we have $(E_{3}+E_{4}-E'_{3}-E'_{4})\sim p_{T}^{2}/\sqrt{s}\sim{p'}_{T}^{2}/\sqrt{s}$.
If $p_{T}^{2}|t|/\sqrt{s}\leqslant p_{T}^{2}\tau/\sqrt{s}\ll1$ or
$\tau\ll\sqrt{s}/p_{T}^{2}$, we can expand the phase factor in Eq.
(\ref{eq:approx4}) as \begin{eqnarray}
e^{i(E_{3}+E_{4}-E'_{3}-E'_{4})t} & \approx & 1+i(E_{3}+E_{4}-E'_{3}-E'_{4})t-\frac{1}{2}(E_{3}+E_{4}-E'_{3}-E'_{4})^{2}t^{2}+\cdots.\label{eq:expansion}\end{eqnarray}
To the leading order, we have $E_{3}\approx E'_{3}$, $E_{4}\approx E'_{4}$
and $e^{i(E_{3}+E_{4}-E'_{3}-E'_{4})t}\approx1$, Eq. (\ref{eq:approx4})
becomes \begin{eqnarray}
\frac{d^{2}\sigma^{(0)}}{d^{2}\mathbf{x}_{T}} & = & \frac{1}{64(2\pi)^{4}|P_{1}\cdot P_{2}|}\int d^{2}\mathbf{p}_{T}d^{2}\mathbf{p}'_{T}e^{-i(\mathbf{p}_{T}-\mathbf{p}'_{T})\cdot\mathbf{x}_{T}}\sum_{i=1,2}\sum_{\lambda_{1},\lambda_{2},\lambda_{4}}\frac{1}{\left|E_{3}p_{4z}^{(i)}-E_{4}p_{3z}^{(i)}\right|}\nonumber \\
 &  & \times M(P_{3}\lambda_{3},P_{4}\lambda_{4})M^{*}(P'_{3}\lambda_{3},P'_{4}\lambda_{4}).\label{eq:dsigma1}\end{eqnarray}
It is interesting to see the above is unique to the leading order
since \begin{equation}
\frac{1}{\left|E_{3}p_{4z}^{(i)}-E_{4}p_{3z}^{(i)}\right|}\approx\frac{1}{\left|E_{3}p_{4z}^{(i)}-E_{4}p_{3z}^{(i)}\right|^{a_{1}}}\frac{1}{\left|E'_{3}{p'}_{4z}^{(i)}-E'_{4}{p'}_{3z}^{(i)}\right|^{a_{2}}}.\end{equation}
The next-to-leading order non-vanishing contribution comes from the
term $\sim t^{2}$ in Eq. (\ref{eq:expansion}) since the linear term
is odd in $t$ whose integral gives zero in the range $[-\tau/2,\tau/2]$.
It reads \begin{eqnarray}
\frac{d^{2}\sigma^{(1)}}{d^{2}\mathbf{x}_{T}} & = & -\frac{\tau^{2}}{1536(2\pi)^{4}|P_{1}\cdot P_{2}|}\int d^{2}\mathbf{p}_{T}d^{2}\mathbf{p}'_{T}e^{-i(\mathbf{p}_{T}-\mathbf{p}'_{T})\cdot\mathbf{x}_{T}}\nonumber \\
 &  & \times\sum_{i=1,2}\sum_{\lambda_{1},\lambda_{2},\lambda_{4}}\frac{(E_{3}+E_{4}-E'_{3}-E'_{4})^{2}}{\sqrt{E_{3}E_{4}E'_{3}E'_{4}}}\frac{E_{3}E_{4}}{\left|E_{3}p_{4z}^{(i)}-E_{4}p_{3z}^{(i)}\right|}\nonumber \\
 &  & \times M(P_{3}\lambda_{3},P_{4}\lambda_{4})M^{*}(P'_{3}\lambda_{3},P'_{4}\lambda_{4}).\label{eq:dsigma2}\end{eqnarray}

For a simple case in the center-of-mass frame of two colliding quarks,
where $p_{1z}=-p_{2z}$, $p_{3z}=-p_{4z}$ and $E_{1}=E_{2}=E_{3}=E_{4}=\sqrt{s}/2$.
To the leading order, we have and $E'_{3}=E'_{4}\approx\sqrt{s}/2$,
then Eq. (\ref{eq:dsigma1}) is evaluated as \begin{eqnarray}
\frac{d^{2}\sigma^{(0)}}{d^{2}\mathbf{x}_{T}} & \approx & \frac{\alpha_{s}^{2}}{36\pi^{2}}\int d^{2}\mathbf{p}_{T}d^{2}\mathbf{p}'_{T}e^{-i(\mathbf{p}_{T}-\mathbf{p}'_{T})\cdot\mathbf{x}_{T}}\nonumber \\
 &  & \times\frac{1}{p_{T}^{2}+\mu_{m}^{2}}\frac{1}{{p'}_{T}^{2}+\mu_{m}^{2}}\left\{ 1+i\lambda_{3}\mathbf{n}_{1}\cdot\left[\mathbf{n}\times\frac{\mathbf{p}_{T}-\mathbf{p}'_{T}}{\sqrt{s}}\right]\right\} \nonumber \\
 & = & \frac{\alpha_{s}^{2}}{9}\left[A^{2}(x_{T})+\frac{2}{\sqrt{s}}\lambda_{3}\mathbf{n}\cdot(\mathbf{n}_{1}\times\widehat{\mathbf{x}}_{T})A(x_{T})\frac{dA(x_{T})}{dx_{T}}\right]\nonumber \\
 & = & \frac{d^{2}\sigma_{upol}^{(0)}}{d^{2}\mathbf{x}_{T}}+\lambda_{3}\frac{d^{2}\sigma_{pol}^{(0)}}{d^{2}\mathbf{x}_{T}},\end{eqnarray}
where we have taken only the magnetic gluon exchange into account
and the magnetic mass $\mu_{m}$ is introduced to regulate the divergence
arising from the soft gluon exchange. We denoted $\mathbf{n}_{1}=\mathbf{e}_{z}$
as the direction of $\mathbf{p}_{1}$ and used \begin{eqnarray}
A(x_{T}) & = & \int d^{2}\mathbf{p}_{T}e^{\pm i\mathbf{p}_{T}\cdot\mathbf{x}_{T}}\frac{1}{p_{T}^{2}+\mu_{m}^{2}}=2\pi\int_{0}^{p_{T}^{cut}}dp_{T}\frac{p_{T}J_{0}(p_{T}x_{T})}{p_{T}^{2}+\mu_{m}^{2}},\end{eqnarray}
where $J_{0}$ is the Bessel function and $p_{T}^{cut}$ is the cutoff
to regulate the ultraviolet divergence. The polarized and unpolarized
differential cross sections can be obtained \begin{eqnarray}
\frac{d^{2}\sigma_{upol}^{(0)}}{d^{2}\mathbf{x}_{T}} & = & \frac{\alpha_{s}^{2}}{9}A^{2}(x_{T}),\nonumber \\
\frac{d^{2}\sigma_{pol}^{(0)}}{d^{2}\mathbf{x}_{T}} & = & \frac{2\alpha_{s}^{2}}{9\sqrt{s}}\mathbf{n}\cdot(\mathbf{n}_{1}\times\widehat{\mathbf{x}}_{T})A(x_{T})\frac{dA(x_{T})}{dx_{T}}.\label{eq:zero-order}\end{eqnarray}
Note that the above result is just Eqs. (40,41) in Ref. \cite{Gao:2007bc}. 

The next-to-leading order contribution is evaluated as \begin{eqnarray}
\frac{d^{2}\sigma^{(1)}}{d^{2}\mathbf{x}_{T}} & \approx & -\frac{\tau^{2}\alpha_{s}^{2}}{216\pi^{2}s}\int d^{2}\mathbf{p}_{T}d^{2}\mathbf{p}'_{T}e^{-i(\mathbf{p}_{T}-\mathbf{p}'_{T})\cdot\mathbf{x}_{T}}(p_{T}^{2}-{p'}_{T}^{2})^{2}\nonumber \\
 &  & \times\frac{1}{p_{T}^{2}+\mu_{m}^{2}}\frac{1}{{p'}_{T}^{2}+\mu_{m}^{2}}\left\{ 1+i\lambda_{3}\mathbf{n}_{1}\cdot\left[\mathbf{n}\times\frac{\mathbf{p}_{T}-\mathbf{p}'_{T}}{\sqrt{s}}\right]\right\} \nonumber \\
 & = & \frac{d^{2}\sigma_{upol}^{(1)}}{d^{2}\mathbf{x}_{T}}+\lambda_{3}\frac{d^{2}\sigma_{pol}^{(1)}}{d^{2}\mathbf{x}_{T}},\end{eqnarray}
where we have used \begin{eqnarray}
(E_{3}+E_{4}-E'_{3}-E'_{4})^{2} & = & \left[\sqrt{s}-2\sqrt{s/4-p_{T}^{2}+{p'}_{T}^{2}})\right]^{2}\approx\frac{(p_{T}^{2}-{p'}_{T}^{2})^{2}}{s/4}.\end{eqnarray}
The unpolarized part reads \begin{eqnarray}
\frac{d^{2}\sigma_{upol}^{(1)}}{d^{2}\mathbf{x}_{T}} & = & -\frac{\tau^{2}\alpha_{s}^{2}}{216\pi^{2}s}\int d^{2}\mathbf{p}_{T}d^{2}\mathbf{p}'_{T}e^{-i(\mathbf{p}_{T}-\mathbf{p}'_{T})\cdot\mathbf{x}_{T}}(p_{T}^{4}+{p'}_{T}^{4}-2p_{T}^{2}{p'}_{T}^{2})\nonumber \\
 &  & \times\frac{1}{p_{T}^{2}+\mu_{m}^{2}}\frac{1}{{p'}_{T}^{2}+\mu_{m}^{2}}=-\frac{\tau^{2}\alpha_{s}^{2}}{27s}(A_{0}A_{1}-A_{2}^{2}),\label{eq:first-order1}\end{eqnarray}
where we used \begin{eqnarray}
A_{i}(x_{T}) & \equiv & \int_{0}^{p_{T}^{cut}}dp_{T}\frac{p_{T}^{n_{i}}}{p_{T}^{2}+\mu_{m}^{2}}J_{0}(p_{T}x_{T}).\end{eqnarray}
For $i=0,1,2$ with $n_{i}=1,5,3$. The polarized part turns out to
be \begin{eqnarray}
\frac{d^{2}\sigma_{pol}^{(1)}}{d^{2}\mathbf{x}_{T}} & = & -\frac{\tau^{2}\alpha_{s}^{2}}{27s^{3/2}}\mathbf{n}\cdot(\mathbf{n}_{1}\times\widehat{\mathbf{x}}_{T})\frac{d}{dx_{T}}(A_{0}A_{1}-A_{2}^{2}).\label{eq:first-order2}\end{eqnarray}

\begin{figure}
\caption{\label{fig:polarized-crs}The polarized and unpolarized differential
cross sections in the leading order from Eq. (\ref{eq:zero-order})
(left panel) and the next-to-leading order from Eqs. (\ref{eq:first-order1},\ref{eq:first-order2})
(right panel). The parameters are set to the same values as in Fig.
\ref{fig:exact-result}. }

\vspace{0.2cm}

\includegraphics[scale=0.5]{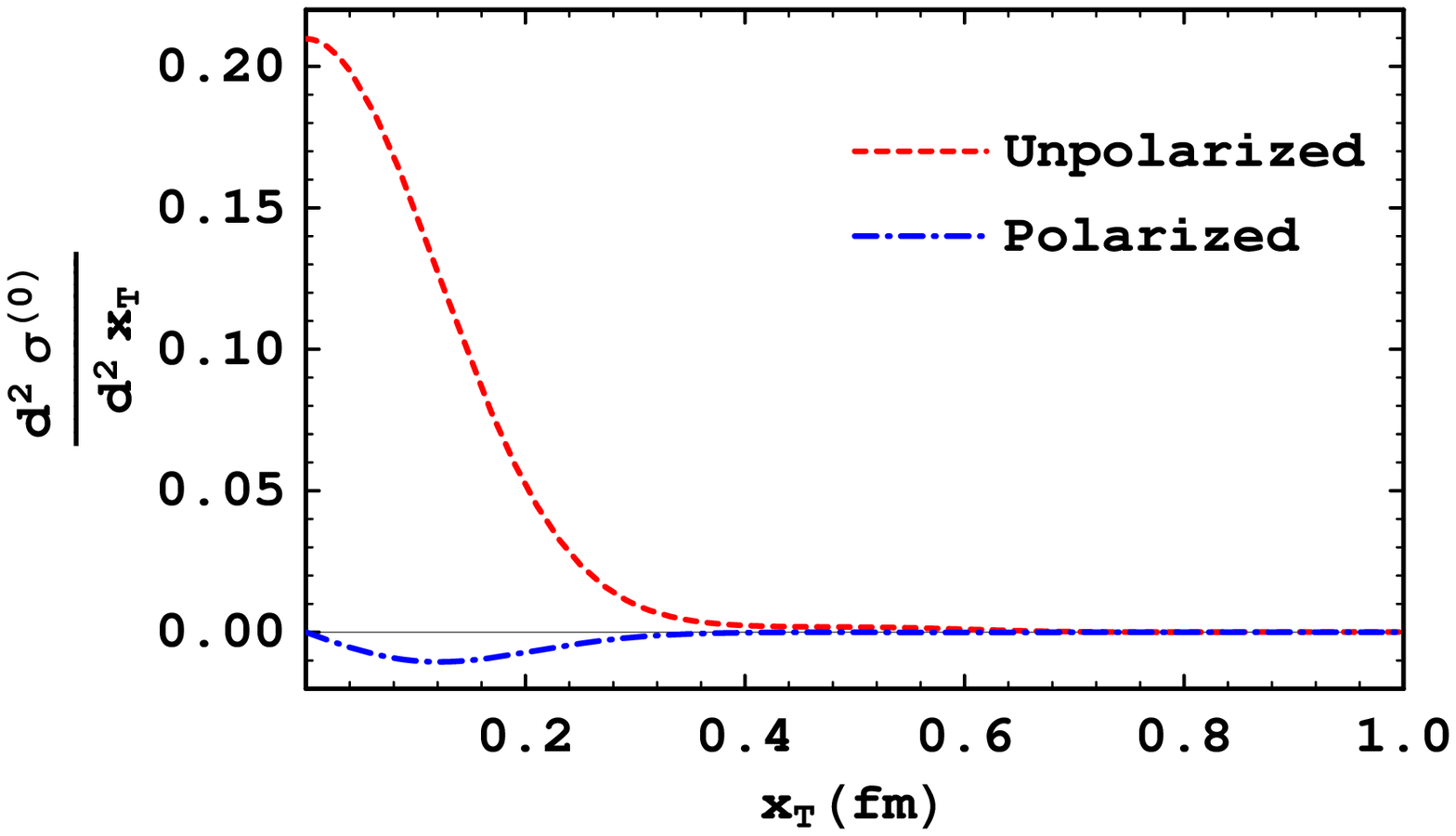}
\hspace{0.5cm}\includegraphics[scale=0.52]{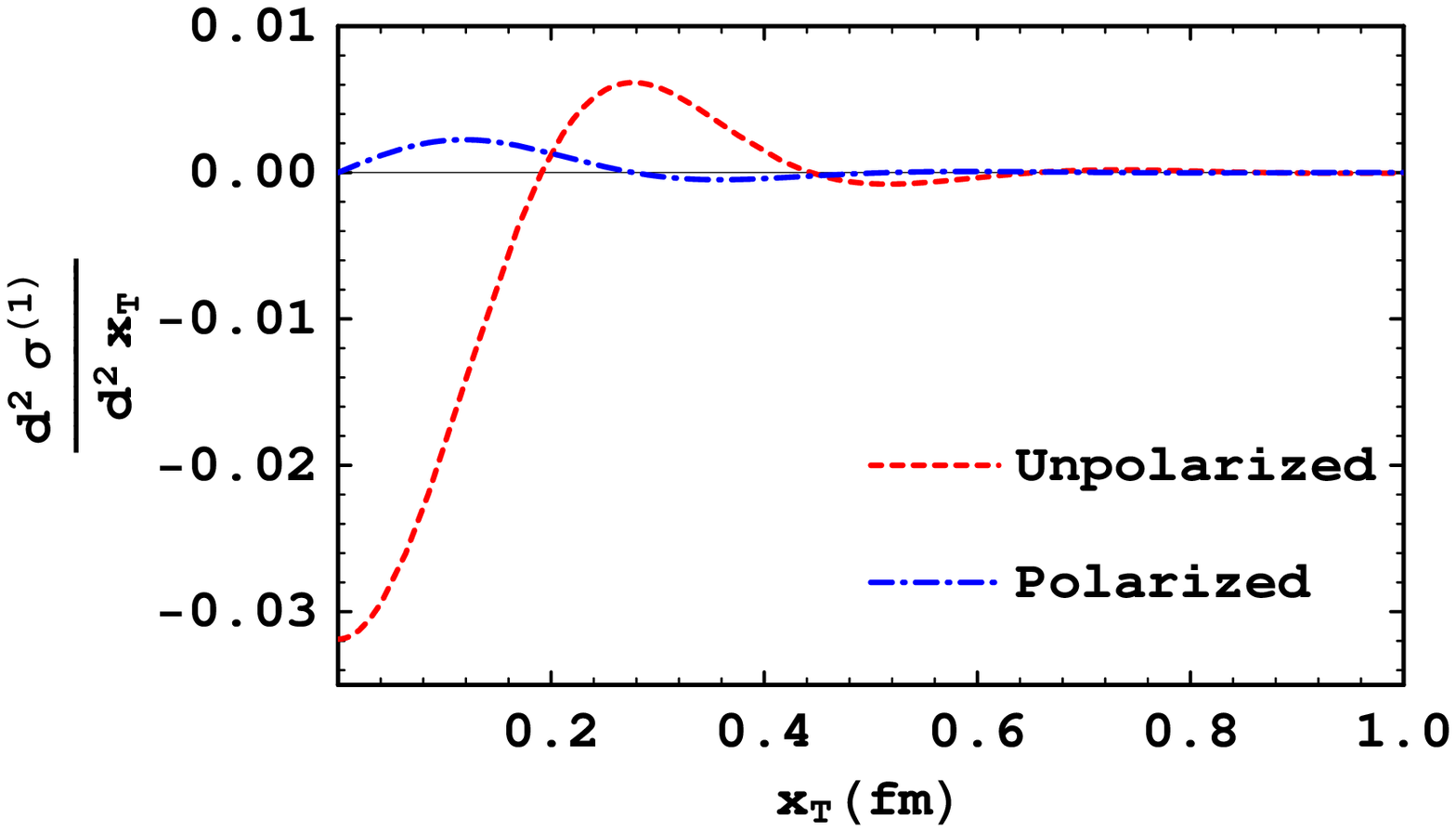}
\end{figure}

The numerical results of Eqs. (\ref{eq:zero-order},\ref{eq:first-order1},\ref{eq:first-order2})
are shown in Fig. \ref{fig:polarized-crs}. We see that both the leading
and next-to-leading parts are damped out above 0.4 fm. The next-to-leading
contributions of the unpolarized and polarized parts are about 1/6
of the leading counterparts. Both the unpolarized and polarized differential
cross sections show an oscillation feature. The sums of the leading
and next-to-leading contributions are shown as the red dotted (unpolarized)
and blue dash-dotted line (polarized) in Fig. \ref{fig:exact-result}.
We already mentioned that a cutoff in transverse momentum $p_{t}^{cut}$
is needed to regulate the integrals in the leading and next-to-leading
differential cross sections. The cross section results depend on $p_{t}^{cut}$
in the perturbation. The time scale $\tau$ is also a quantity to
be determined. We can find the range of the $\tau$ by equating the
exact result $d^{2}\sigma/d^{2}\mathbf{x}_{T}$ from Eq. (\ref{eq:cross-section1})
and the sum $d^{2}(\sigma^{(0)}+\sigma^{(1)})/d^{2}\mathbf{x}_{T}$
in the perturbation approach from Eqs. (\ref{eq:zero-order},\ref{eq:first-order1},\ref{eq:first-order2})
at a specified value of $p_{t}^{cut}$. 

\section{Summary and discussion}

We have proposed a general approach to the differential cross section
with respect to impact parameters, which is well-defined and free
of ambiguity existing in the conventional approach. The main difference
of our approach from the conventional one is that (1) in the conventional
approach the transfer of small transverse momenta is implied, while
the general approach is valid for all transverse momenta; (2) there
are two independent delta functions in the general approach for energy
conservation in the cross section formula, which arises from fixing
impact parameters in the final state partons making the total final
state energy be uncertain. While in the conventional approach the
two delta functions for energy conservation are identical turning
the second one to be the infinite interaction time $\tau$. 

As a simple illustration of our formalism, we evaluated the polarized
and unpolarized differential cross sections at small angle quark-quark
scatterings in the center-of- mass system of the colliding quarks.
To smoothly connect the general approach and the conventional one,
we propose an expansion in terms of $\Delta E=E_{f}-E_{i}\sim1/\tau$
with $E_{i}$ and $E_{f}$ the initial and final state energies in
collisions. The leading order contribution reproduced the conventional
result, i.e. that of Ref. \cite{Gao:2007bc}. 

The general formulation in this paper can also be applied to many
other parton-parton scatterings in heavy ion collisions or even proton-proton
collisions \cite{Frankfurt:2003td}. 

{\bf Acknowledgement}. The authors thank Zuo-tang Liang and Xin-nian
Wang for insightful discussions. Q.W. acknowledges supports in part
by the startup grant from University of Science and Technology of
China (USTC) in association with \textit{100-talent} project of Chinese
Academy of Sciences (CAS) and by NSFC grant No. 10675109.

\end{document}